\pdfoutput=1

\documentclass[11pt]{article}
\usepackage{subcaption}

\usepackage[final]{acl}

\usepackage{times}
\usepackage{latexsym}
\usepackage{amsmath}
\usepackage{amssymb}
\usepackage{tabularx}
\usepackage{array}
\usepackage{makecell} 
\usepackage{booktabs} 

\usepackage[T1]{fontenc}

\usepackage[utf8]{inputenc}

\usepackage{microtype}

\usepackage{inconsolata}

\usepackage{graphicx}

\title{Lyrics Matter: Exploiting the Power of Learnt Representations for Music Popularity Prediction}

\author{
  Yash Choudhary\textsuperscript{1} \quad
  Preeti Rao\textsuperscript{2} \quad
  Pushpak Bhattacharyya\textsuperscript{3} \\
  \textsuperscript{1}Indian Institute of Technology Bombay, Mumbai, India \\
  \textsuperscript{2}Department of Electrical Engineering, IIT Bombay, India \\
  \textsuperscript{3}CFILT, Department of Computer Science, IIT Bombay, India \\
  \texttt{200100173@iitb.ac.in} \quad
  \texttt{prao@ee.iitb.ac.in} \quad
  \texttt{pb@cse.iitb.ac.in}
}

\begin{document}
\maketitle
\begin{abstract}

Accurately predicting music popularity is a critical challenge in the music industry, offering benefits to artists, producers, and streaming platforms. Prior research has largely focused on audio features, social metadata, or model architectures. This work addresses the under-explored role of lyrics in predicting popularity. We present an automated pipeline that uses LLM to extract high-dimensional lyric embeddings, capturing semantic, syntactic, and sequential information. These features are integrated into HitMusicLyricNet, a multimodal architecture that combines audio, lyrics, and social metadata for popularity score prediction in the range 0-100. Our method outperforms existing baselines on the SpotGenTrack dataset, which contains over 100,000 tracks, achieving 9\% and 20\% improvements in MAE and MSE, respectively. Ablation confirms that gains arise from our LLM-driven lyrics feature pipeline (LyricsAENet), underscoring the value of dense lyric representations.
\end{abstract}

\section{Introduction}

In 2023, the global recorded music market generated \$28.6 billion\footnote{\href{https://www.ifpi.org/our-industry/industry-data/}{IFPI Report '23}} in revenue. With the advent of social media and streaming services, defining a single metric for music success has become increasingly challenging \citep{Cosimato2019TheCO, Lee2020DisentangledMM}. Music popularity prediction can help the industry and artists forecast and optimize the potential success of newly composed songs.

Research in music popularity prediction has been driven by the advancements in machine learning with researchers applying classical ML approaches to predict popularity using acoustic features, and further with the growth of social networks, information about music consumers’ tastes capturing consumer response and their evolving music preferences \cite{Seufitelli2023HitSS}. Advancements in deep learning further sharpen the prediction model capability of capturing and learning complex patterns of evolving music taste, and researchers have worked on incorporating multiple modalities such as audio, lyrics and social metadata to predict song success \citep{Zangerle2019HitSP,MartnGutirrez2020AME}. In all these works, the popularity score is typically defined as the time the song remains on the Billboard Top charts, and the evaluation metrics used include MAE, MSE, R\textsuperscript{2} for regression, and accuracy, precision, recall, and F1 for classification. Recent developments in large language models have led to further research in music-related fields such as recommendation systems, sentiment/emotion analysis, data augmentation, understanding and composing song lyrics, using song lyrics text as the data source \citep{10.1145/3607827.3616842,Sable2024EnhancingMM,ma2024foundation,ding2024songcomposer}. Music Popularity Prediction research has still not fully exploited the power of lyrics in the models, while recent research have shown lyrics contributing significantly to song popularity \citep{Yu_Cheung_Ahn_Dhillon_2023}. Through our work, we address the gap in the existing literature with the following main contributions:
\begin{enumerate}
    \item A novel automated lyric feature extraction pipeline that uses LLMs to encode music lyrics into rich, learned representations. Details discussed in \ref{subsubsec:LyricsAENe}
    
    \item An end to end multimodal deep learning architecture which predicts the popularity score in range (1,100) and outperformed current baseline by 9\% and 20\% in MAE and MSE metrics respectively. Details discussed in \ref{subsec:hitmusiclyricnet}
\end{enumerate}

The next section reviews related work. This is followed by a discussion of our methods, the dataset and our experiments. 

\section{Related Work}
\textbf{Music Popularity Prediction}. Studied as a classification or regression problem in a supervised learning fashion, where a model learns to predict either binary class labels (hit or no-hit) or generate a continuous popularity score \cite{Seufitelli2023HitSS}. These predictions are derived using the song's internal characteristics (audio and lyrics) and associated social factors like artist, genre, user demographics, etc. Song popularity is primarily measured using charts like Billboard \footnote{\href{http://www.billboard.com/charts/hot-100}{Billboard Hot 100}} and UK Singles Charts\footnote{\href{https://www.officialcharts.com/charts/singles-chart/}{Official Singles Chart Top 100}} \citep{Bischoff2009,Askin2017,Kim2014,LeeLee2018}
, which rank songs based on sales, radio airplay, and streaming activity. Researchers determine success metrics based on these rankings, time on top charts, and other measures, including economic metrics like merchandise sales and user engagement metrics on social media and streaming services \cite{Seufitelli2023HitSS}.

Traditional research focused on using various machine learning techniques, including Logistic Regression, Decision Trees, Support Vector Machines (SVM), Bayesian Networks, Naive Bayes, Random Forest Ensemble, XGBoost, and K-Nearest Neighbors (KNN). These approaches advanced further to neural networks and deep learning techniques, building much stronger predictive models. A significant number of studies \citep{10.1007/978-3-642-03348-3_8, Herremans03072014,Zangerle2019HitSP,10.1145/3539637.3556993} focused on using acoustic characteristics of songs along with metadata that includes factors such as social influences. Other works such as \cite{Dhanaraj2005AutomaticPO, singhi2015songlyrics, MartnGutirrez2020AME} also emphasized the importance of song lyrics in determining song success using handcrafted text-based features that captured sentiment, emotions, and the syntactic structure of lyrics. These studies were often limited by their capabilities to capture central expressions of the song's lyrics.

Multiple datasets have been released to drive research further and quench the thirst of data-heavy deep learning models. This includes Million Song Dataset\footnote{\href{http://millionsongdataset.com/}{Million Song Dataset}}, SpotGenTrack\footnote{\href{https://data.mendeley.com/datasets/4m2x4zngny/1}{SpotGenTrack}}, and AcousticBrainz \footnote{\href{http://acousticbrainz.org/}{AcousticBrainz}} sourced from different platforms like Spotify, Billboard, Genius \footnote{\href{https://genius.com/}{Genius.com}}, Youtube, and others. These datasets comprise a wide range of features, from low-level features like Mel-Frequency Cepstral Coefficients (MFCCs), lyrics text, and temporal features to high-level audio features such as danceability and loudness. Additionally, they include metadata on artists, albums, genres, demographics, and other relevant information.

\textbf{Learned Representations of Lyrics}. 
Lyrics form an integral part of music and carry a deep emotional meaning, which can strongly influence how listeners feel\textemdash sometimes even more than the song’s acoustic features alone \cite{SinghiBrown2015}. Yet, lyrics have often been overlooked as compared to acoustic attributes and social metrics of songs \cite{Seufitelli2023HitSS}. Earlier studies used methods like Probabilistic Latent Semantic Analysis (PLSA) \cite{10.1145/312624.312649} to capture the semantic content of lyrics, which helped researchers understand their role in defining a “hit” song \cite{Dhanaraj2005AutomaticPO}. Later work moved beyond basic semantic analysis, focusing on more detailed features. For instance,  \cite{Hirjee2010} and \cite{Singhi2014} relied on various rhyme and syllable characteristics to predict hit songs using only their lyrics, while other researchers applied Latent Dirichlet Allocation (LDA) \cite{Blei2003} to discover thematic topics within lyrics \cite{10.1145/2872518.2889402}.

Progress of deep learning techniques advanced the use of multimodal approaches that combine lyrics with audio and metadata, using stylometric analysis to extract lyric text features \cite{MartnGutirrez2020AME}. Sentiment analysis also emerged as a way to glean emotional insights from lyrics when predicting popularity \cite{9190613}. More recent research has turned to learned lyric representations, such as embeddings \cite{kamal2021songpopularity,lyric-document-embeddings}, which offer a more robust way to capture lyrical meaning. \cite{10.1007/978-3-030-15719-7_4} demonstrated that these distributed representations can effectively predict both genre and popularity, reducing the need for handcrafted features. Datasets such as Music4All-Onion \cite{10.1145/3511808.3557656} provide lyric embeddings that make it easier to study how lyrical content relates to a song’s success. Finally, a recent study found that a song’s lyrical uniqueness has a significant contribution towards its popularity \cite{Yu_Cheung_Ahn_Dhillon_2023}, using TF–IDF for lyric vector representation; however, this approach inherently lacks the capacity to capture deeper sequential and contextual nuances, emphasizing the growing importance of learning robust, richer representations of lyrics to better understand what makes certain songs resonate with audiences.


To the best of our knowledge, there are limitations in existing literature for efficient automated lyrics feature extraction that are expressive and capture the underlying complexity of song lyrics. Thus, we have built a novel pipeline to exploit the power of Large Language Models. It has the potential to provide rich lyric representations that encapsulate both semantic and syntactic understanding, while preserving the sequential structure of the lyrics.

\section{Methodology}\label{sec:Methodology}
In this section, we provide the theoretical foundation of our approach. We begin by defining the problem of music popularity prediction in mathematical equations. This is followed by explaining the baseline approach and its implementation, including details of the dataset. Finally, we present a formal description of our proposed architecture.

\subsection{Problem Formulation}
Given a song \( S \), its features are represented in a multi-dimensional space \( X \in \mathbb{R}^d \), which comprises three key modalities: audio waveform \( w \in \mathbb{R}^k \), lyrical text \( l \in \mathbb{R}^n \), and metadata attributes \( m \in \mathbb{R}^p \), where \( d = k + n + p \) represents the total dimensionality of our feature space. Our primary objective is to extract meaningful features from the song lyrics to effectively encode each song into a unique vector representation. Next, the prediction task is formulated as learning a mapping function \(f: X \rightarrow Y,\) where we minimize the expected prediction error: \(\mathbb{E}[(f(X) - Y)^2]\) across the training distribution. Here, \( Y \in \mathbb{R} \) represents the continuous popularity score.

\subsection{Baseline Methodology}\label{subsec:baseline-methodology}

\begin{figure}[h]
  \includegraphics[width=0.98\linewidth]{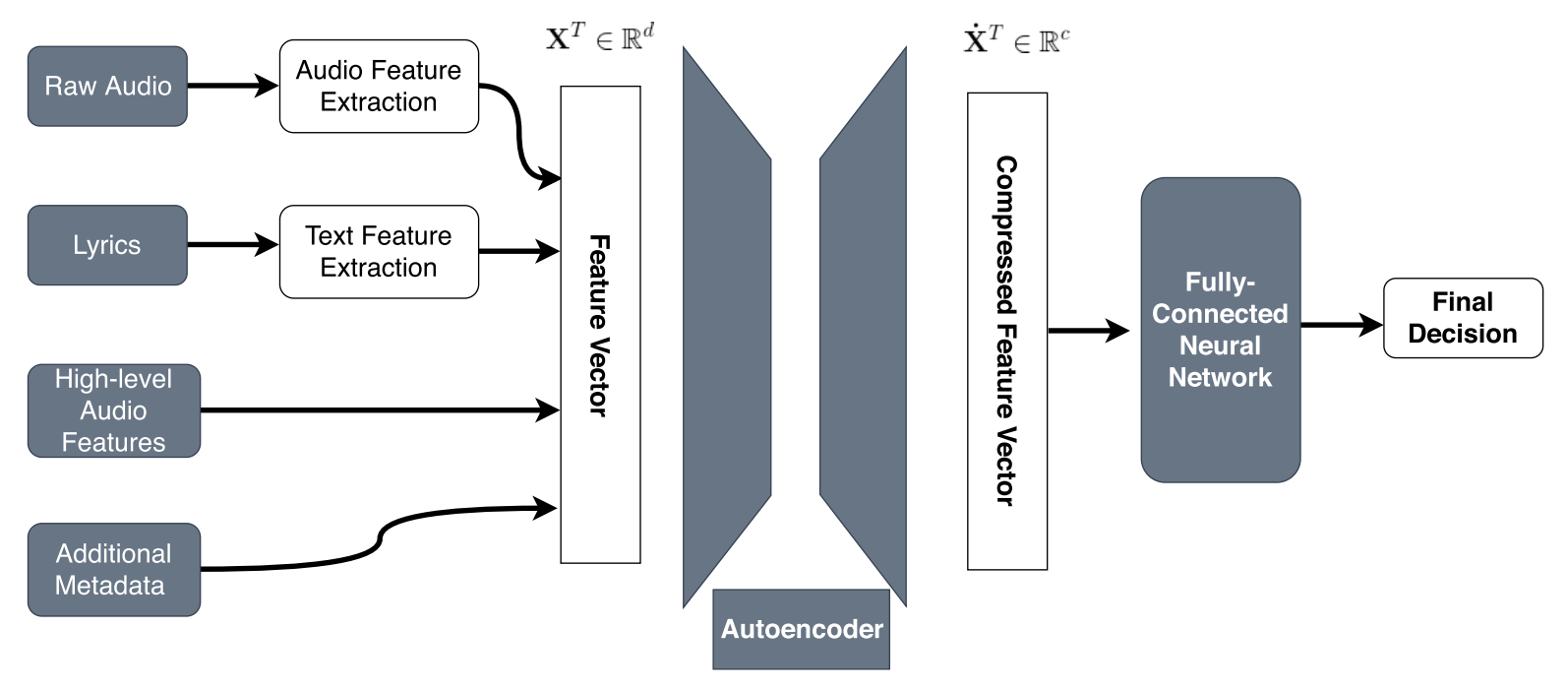} \hfill
  \caption {\label{fig:hitmusicnet} Diagram of the HitMusicNet pipeline outlining the principal functionalities and data components. Image src \cite{MartnGutirrez2020AME}.}
  \vspace{-0.1cm}

\end{figure}

We trained \textit{HitMusicNet}, a multimodal end-to-end Deep Learning architecture as proposed by \cite{MartnGutirrez2020AME} and validated the results using the SpotGenTrack Popularity Dataset (SPD). The model outputs a popularity score between 1 and 100, using audio features, text features, and metadata containing artist and demographic information as inputs. A complete description of the feature set used is provided in Table~\ref{tab:feature-summary}. 

\begin{table}[h]
\centering
\small 
\begin{tabularx}{\linewidth}{lX}
\hline
\textbf{Feature Type}        & \textbf{Features} \\
\hline
\textbf{Text Features}& Sentence count, Avg words, Word count, Avg syllables/word, Sentence similarity, Vocabulary wealth \\
\textbf{High-Level Audio}    & Danceability, Energy, Key, Loudness, Mode, Speechiness, Acousticness, Instrumentalness, Liveness, Valence, Tempo, Duration, Time Signature \\
\textbf{Low-Level Audio}     & Mel-spectrogram, MFCCs, Tonnetz, Chromagram, Spectral Contrast, Centroid, Bandwidth, Zero-Crossing Rate \\
\textbf{Meta-Data Features}  & Artist followers, Artist popularity, Available markets \\
\hline
\end{tabularx}
\caption{\label{feature-summary}
\label{tab:feature-summary} Summary of features used in the HitMusicNet architecture \cite{MartnGutirrez2020AME}.
}
\end{table}

\textit{HitMusicNet} architecture as shown in Fig~\ref{fig:hitmusicnet}, employs an autoencoder for feature compression through two encoder layers with dimensions $d/2$ and $d/3$, followed by a bottleneck layer of $d/5$. Each layer uses ReLU activation, and the output layer employs a sigmoid activation for reconstruction. The autoencoder was trained using the Adam optimizer and an MSE loss function. The compressed features are then passed through a fully connected neural network with four layers, where the number of neurons in each layer is scaled by factors $\alpha=1$, $\beta=1/2$, and $\gamma=1/4$. The model is trained using an 80\%-20\% train-test split with stratified cross-validation (SCV) using $k=5$. These settings helped us in effectively replicating the baseline results on the SPD dataset. 

\subsection{Dataset}

\begin{figure}[h!]
\includegraphics[width=0.98\linewidth]{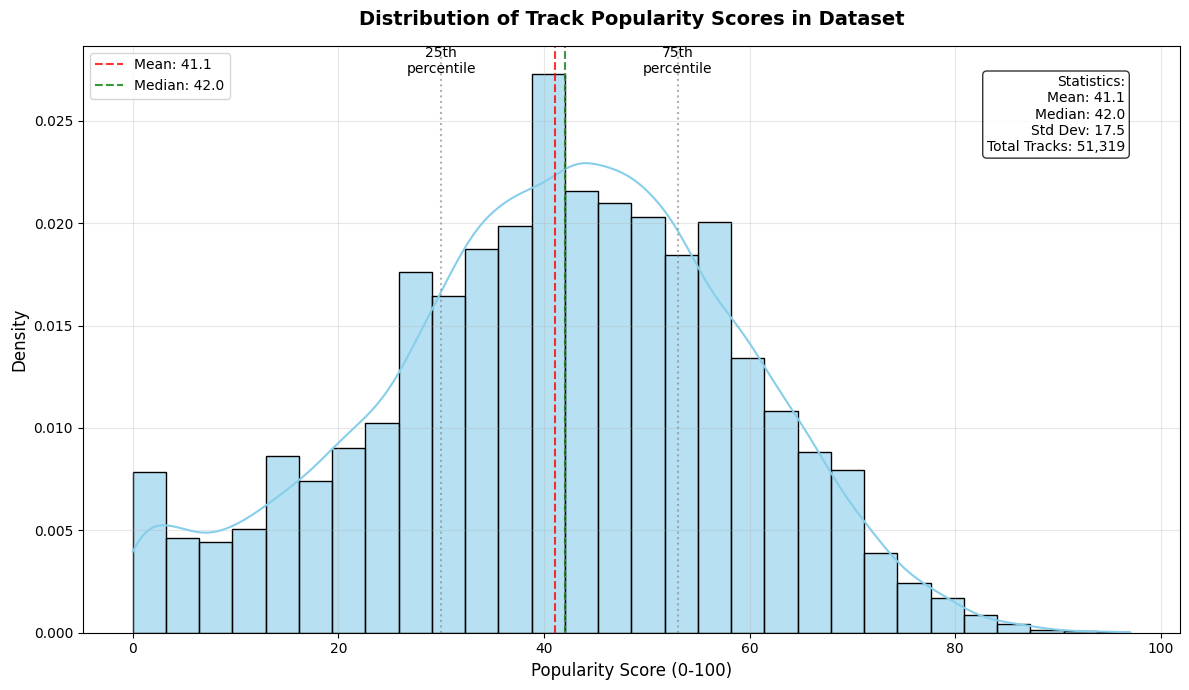} \hfill
  \caption {\label{fig:SPDcleanedist} Popularity Distribution in cleaned SpotGenTrack(SPD) with $\mu = 41.11$ and a standard deviation of $\sigma = 17.51$.} 
  \vspace{-0.45cm}
\end{figure} 

We use the SpotGenTrack Popularity Dataset (SPD), originally introduced by \citet{MartnGutirrez2020AME}, which contains 101,939 tracks from 56,129 artists and 75,511 albums. Tracks are sourced from Spotify and Genius APIs, covering the top 50 playlists across 26 countries. Spotify provides track-level popularity scores ranging from 1 to 100. These scores follow a Gaussian distribution with $\mu = 40.02$ and $\sigma = 16.79$. The dataset includes low-level audio features extracted from raw waveforms, high-level audio descriptors, stylometric text features derived from lyrics, and metadata such as artist popularity and market reach. To ensure data quality, we applied filtering steps to remove noisy lyric entries. Specifically, we excluded tracks with lyrics shorter than 100 or longer than 7,000 characters, which often contained placeholders or irrelevant content. Additionally, we restricted the dataset to five major languages: English, Spanish, Portuguese, French, and German—discarding other languages that constituted less than 1\% of the data. This resulted in a cleaned corpus of 74,206 tracks, comprising 51,319 in English and 22,887 in the remaining languages. The cleaned popularity distribution as shown in Fig~\ref{fig:SPDcleanedist} maintained the original characteristics, with $\mu = 41.11$ and $\sigma = 17.51$, ensuring that no sampling bias was introduced. We min-max normalize Spotify popularity scores from [0,100] to [0,1] for training stability. All metrics are reported on the normalized scale.

We further considered multiple open-source music popularity datasets for benchmarking HitMusicLyricNet, but none of them meet our multimodal data requirements: the \textsc{TPD} dataset \cite{Karydis2016MusicalTP} lacked lyrical and social metadata; the \textsc{MSD} dataset \cite{BertinMahieux2011TheMS} offered only bag-of-words lyrics; \textsc{HSP-S} and \textsc{HSP-L} datasets \cite{Vtter2021NovelDF} omitted full lyrical content; the \textsc{MusicOSet} \cite{mariana_o_silva_2019_4904639} included lyrics but lacked detailed audio-level features. Further, the \textsc{LFM-2b} dataset \cite{10.1145/3498366.3505791}, has copyright issues.

\subsection{HitMusicLyricNet}\label{subsec:hitmusiclyricnet}
This section details our proposed HitMusicLyricNet, an end-to-end multimodal deep learning architecture built upon the foundation of HitMusicNet. HitMusicLyricNet comprises of three key components: AudioAENet, LyricsAENet, and MusicFuseNet. AudioAENet compresses the low-level audio features. LyricsAENet compresses the lyric embeddings into a fixed-size representation using an Autoencoder, thereby encoding information while reducing noise. MusicFuseNet then combines these compressed audio and lyric representations with metadata and high-level audio features as described in Table~\ref{tab:feature-summary}.

In the HitMusicNet architecture, a single auto-encoder compressed the combined feature vector of audio, lyrics, and metadata. We hypothesize that this can lead to information loss, particularly for the less abundant lyrics and metadata features. We believe that lyrics and metadata features should be fed directly into the popularity prediction network to retain their predictive power for song popularity. Furthermore, our reasoning behind the new approach of introducing distinct Autoencoders for audio and lyrics is based on the bipolar and directional nature of lyrics embeddings, requiring a different architecture for compression\cite{balazy-etal-2021-direction}.

\begin{figure*}[t]
  \includegraphics[scale = 0.145]{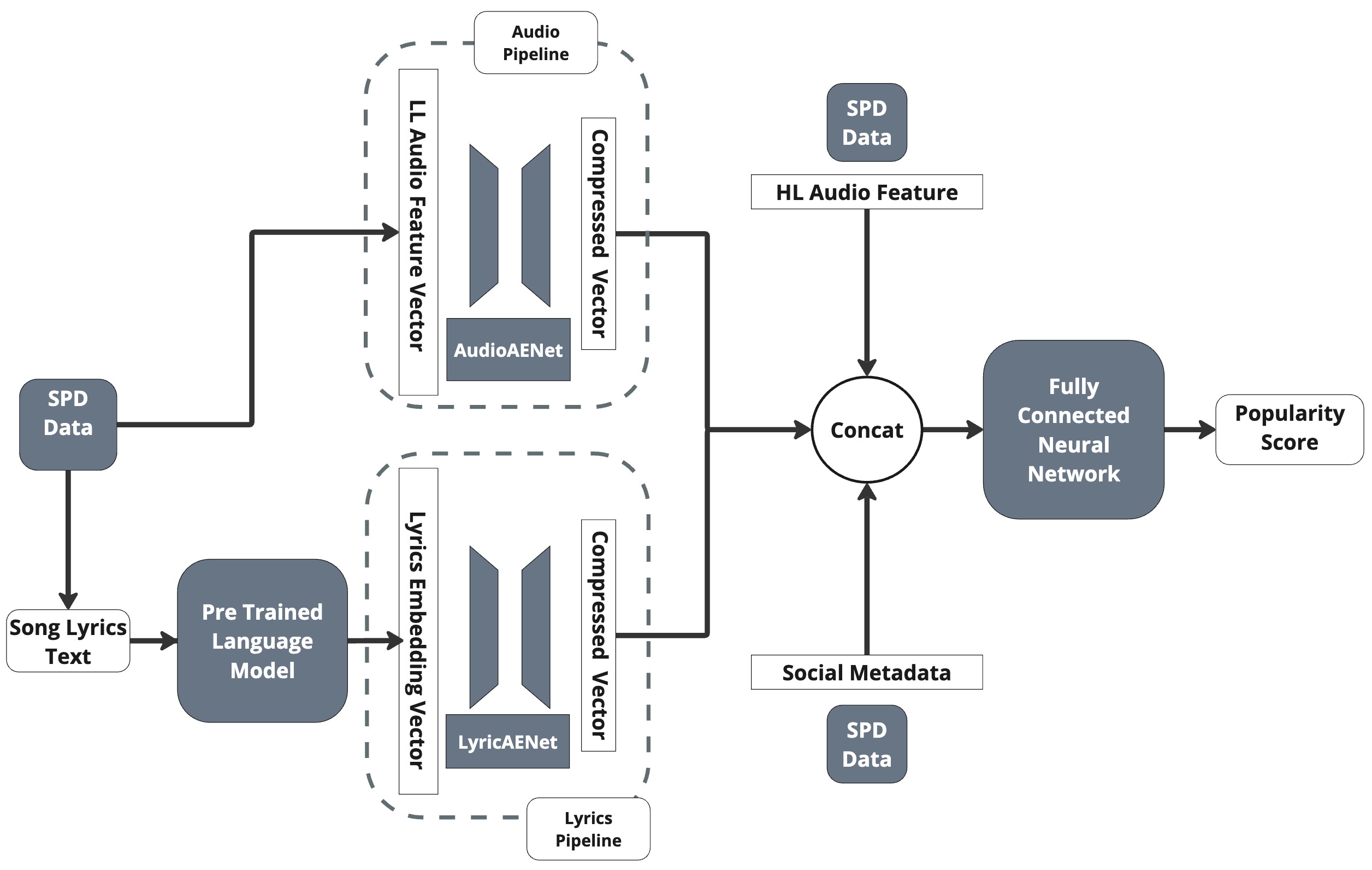} \centering
  \caption {\label{fig:HitMusicLyricNet}Block schematic of the \textit{HitMusicLyricNet} architecture comprising of two Autoencoders and a Fully Connected NN predicting popularity score. 'HL' stands for high-level and 'LL' stands for low-level.}
\vspace{-0.2cm}
\end{figure*}

\subsubsection{AudioAENet}
The Autoencoder used for compression has a similar architecture to that of MusicAENet in \cite{MartnGutirrez2020AME}, but takes in only low-level audio features as described in Table~\ref{tab:feature-summary} for compression. For input dimension d = 209, it gradually compresses the data to dimension d/2, d/3, and d/5. The output layer employs a sigmoid activation for reconstruction, whereas all remaining layers use ReLU activation functions. The model is trained using the Adam optimizer with a MSE loss function, achieving a loss value in the range of 1e-5, indicating negligible loss in compression.

\subsubsection{LyricsAENet}\label{subsubsec:LyricsAENe}
LyricsAENet implements a tied-weights autoencoder architecture \cite{li2018on} designed to reduce parameter size and risk of overfitting. Compressing lyric embeddings is susceptible to overfitting due to high dimensionality. The encoder follows a progressive compression with the following dimensions (d/2, d/4, d/8), followed by bottleneck layers (d/12 or d/16). The decoder mirrors the structure in reverse order, utilizing the transpose of the encoder weight. The progressive dimensional reduction is designed to minimize reconstruction losses in compressed embeddings extracted out of language models and LLMs such as BERT \cite{devlin-etal-2019-bert}, LLaMA 3 Herd \cite{grattafiori2024llama3herdmodels}, and OpenAI's embedding models\footnote{\href{https://platform.openai.com/docs/guides/embeddings}{Open AI text Embedding Model}}. 

 We use Scaled Exponential Linear Unit (SELU) \cite{klambauer2017self} as the activation function for its self-normalizing characteristics and the ability to handle the bipolar nature of embeddings. Comparative analyses include alternate activation functions such as the Sigmoid Linear Unit (SiLU) \cite{elfwing2018sigmoid} and the Gaussian Error Linear Unit (GELU) \cite{hendrycks2016gaussian}. LyricsAENet was trained using the Adam optimizer with a MSE loss function, achieving loss values of approximately 1e-5. To further refine the training process, we incorporated a directional loss function inspired by \cite{balazy-etal-2021-direction} to preserve the directional characteristics of the embeddings during compression. This combined loss function is defined as:
\begin{equation}
\label{eq:Dir Loss}
L(Y, \bar{Y}) = \alpha_1 \cdot \text{MSE}(Y, \bar{Y}) + \alpha_2 \cdot CD(Y, \bar{Y}),
\end{equation}

where $\text{MSE}(Y, \bar{Y})$ represents the Mean Squared Error. $CD(Y, \bar{Y})$ denotes the Cosine Distance, which captures the angular similarity between the vectors $Y$ and $\bar{Y}$. The constants $\alpha_1$ and $\alpha_2$ control the relative importance of the two loss terms.

\begin{table*}[h!]
    \centering
    \begin{minipage}[t]{0.48\textwidth}
        \centering
        \small
        \begin{tabular}{>{\centering\arraybackslash}p{1.9cm}>{\centering\arraybackslash}p{1.3cm}>{\centering\arraybackslash}p{1.3cm}>{\centering\arraybackslash}p{1.3cm}}
            \hline
            \textbf{LyricsAENet Config} & \textbf{MAE (Train)} & \textbf{MAE (Val)} & \textbf{MAE (Test)} \\
            \hline
            \textit{SELU, MSE}  & \textbf{0.0769} & \textbf{0.0746} & \textbf{0.0775} \\
            \textit{SiLU, MSE} & 0.0736 & 0.0731 & 0.0790 \\
            \textit{GELU, MSE} & 0.0740 & 0.0731 & 0.0792 \\
            \textit{SELU, Dir.} & 0.0741 & 0.0740 & 0.0799 \\
            \hline
        \end{tabular}
        \caption{\label{tab:Interesults1} Results of training and testing HitMusicLyricNet on cleaned SPD data with various LyricAENet configurations (activation function, loss function), using BERT Large embeddings throughout. ‘Dir’ indicates directional loss~\ref{eq:Dir Loss}.}
    \end{minipage}%
    \hfill
    \begin{minipage}[t]{0.48\textwidth}
        \centering
        \small
        \begin{tabular}{>{\centering\arraybackslash}p{2cm}>{\centering\arraybackslash}p{1.3cm}>{\centering\arraybackslash}p{1.3cm}>{\centering\arraybackslash}p{1.3cm}}
            \hline
            \textbf{Embeddings Model} & \textbf{MAE (Train)} & \textbf{MAE (Val)} & \textbf{MAE (Test)} \\
            \hline
            \textit{BERT large}  & \textbf{0.0793} & \textbf{0.0784} & \textbf{0.0786} \\
            \textit{Llama 3.1 8B} & 0.0774 & 0.0759 & 0.0795 \\
            \textit{Llama 3.2 1B} & 0.0775 & 0.0754 & 0.0800 \\
            \textit{Llama 3.2 3B} & 0.0781 & 0.0766 & 0.0798 \\
            \textit{OpenAI Small} & 0.0746 & 0.0738 & 0.0788 \\
            \textit{OpenAI Large} & \textbf{0.0761} & \textbf{0.0743} & \textbf{0.0772} \\
            \hline
        \end{tabular}
        \caption{ \label{tab:Interesults2} Results of training and testing HitMusicLyricNet on cleaned SPD data with different lyric embeddings sent to LyricAENet (Selu activation, MSE loss).}
    \end{minipage}
\end{table*}

\begin{table*}[h!]
  \small
  \begin{tabular}{>{\centering\arraybackslash}p{4cm}>{\centering\arraybackslash}p{1.5cm}>{\centering\arraybackslash}p{1.5cm}>{\centering\arraybackslash}p{1.3cm}>{\centering\arraybackslash}p{1.5cm}>{\centering\arraybackslash}p{1.3cm}>{\centering\arraybackslash}p{1.5cm}>{\centering\arraybackslash}p{1.5cm}}
    \hline
    \textbf{Model {Config}} & \textbf{Dataset {Config}} & \textbf{MSE (Train)} & \textbf{MSE (Val)} & \textbf{MAE (Train)} & \textbf{MAE (Val)} & \textbf{MAE (Test)} \\
    \hline
    \textit{HitMusicNet} & SPD & 0.0116 & 0.0115 & 0.0836 & 0.0851 & 0.0862 \\
    \textit{HitMusicNet w/o lyrics} & SPD & 0.0114 & 0.0116 & 0.0843 & 0.0859 & 0.0870 \\
    \textit{HitMusicLyricNet} & *SPD & \textbf{0.0095} & \textbf{0.0091} & \textbf{0.0761} & \textbf{0.0743} & \textbf{0.0772} \\
    \textit{HitMusicLyricNet w/o lyrics} & *SPD & 0.0109 & 0.0113 & 0.0818 & 0.0841 & 0.0852 \\
    \hline
  \end{tabular}
  \caption{\label{tab:results} Performance comparisons with the baseline (HitMusicNet) on  SPD and SPD* data respectively, where SPD* denotes cleaned SPD data. Here, we report the best results from Table ~\ref{tab:Interesults2}.}
  \vspace{-0.2cm}

\end{table*}

\subsubsection{MusicFuseNet}
We employ a similar architecture configuration as MusicPopNet by \cite{MartnGutirrez2020AME} for our MusicFuseNet. It uses a concatenation of compressed audio feature vectors from AudioAENet, compressed lyrics embeddings vectors from LyricsAENet, high-level audio features and metadata as mentioned in Table~\ref{feature-summary}. The output of this neural net is a popularity score in the range [0, 1]. The architecture consists of a fully connected network with scaling parameters of (1, 1/2, 1/3) and ReLU activation functions, followed by a Sigmoid activation in the final layer, as empirically validated by \cite{MartnGutirrez2020AME}. To train the model, we used the Adam optimizer with an MSE loss function and applied dropout regularization to mitigate overfitting.

\section{Experiments and Results}

Using the \textbf{Code}\footnote{\href{https://github.com/dmgutierrez/hitmusicnet}{Github: HitMusicNet}} to implement HitMusicNet and selecting the configuration details described in Section~\ref{subsec:baseline-methodology}, we trained HitMusicNet on the SPD dataset with an 80-20 split. To replicate the results obtained by \cite{MartnGutirrez2020AME}, we employed Stratified Cross-Validation (SCV) with k=5 folds and used MAE and MSE as performance metrics. As Table~\ref{tab:results} shows, we achieved similar results on all performance metrics, validating our training and testing strategy. Further, removing the lyrics text features proposed by \cite{MartnGutirrez2020AME} did not degrade the metrics, so we dropped those features for further experiments.

To train HitMusicLyricNet, we extracted lyric embeddings from language models. For open-source models (BERT, Llama), we downloaded vanilla weights from Hugging Face\footnote{\href{https://huggingface.co/}{Hugging Face}} and loaded its vanilla configuration. We used Nvidia A100 GPU for compute requirements. After tokenizing lyrics, we forward-passed them through each model, extracted the last hidden-layer states, and applied max/mean pooling to obtain fixed-size vectors for our Autoencoder. Specifically for BERT, we considered mean pooling and concat (max + CLS token). To get embeddings from OpenAI text models, we used the API endpoint, costing \$3 for the small model and \$6 for the large. Obtaining embeddings via the OpenAI API took $\sim$5 hours due to rate limits, while the open-source took less than an hour. We then studied LLM model architecture and its training corpus effects on music popularity prediction with BERT, BERT Large, Llama 3.1 8B, Llama 3.2 1B, Llama 3.2 3B, and OpenAI text embeddings (small and large).


After extracting these embeddings, we examined different activation layers (Selu, Silu, Gelu) for embedding compression using LyricsAENet and introduced a directional loss function with \(\alpha_1 = 0.5\) and \(\alpha_2 = \frac{0.5}{5}\) as suggested by \cite{balazy-etal-2021-direction}, alongside our standard MSE loss for LyricsAENet, to see their impact on the HitMusicLyricNet performance metric MAE. As reported in Table~\ref{tab:Interesults1}, using SELU with the MSE loss function in LyricsAENet yielded the least MAE error while training HitMusicLyricNet on popularity prediction. Directional loss produced comparable metrics but not enough improvement to be included further. Other activation functions performed closely, but for simplicity and observing ~1–2\% randomness error, we proceeded with SELU and MSE.

Next, we compressed embeddings for different LLM models. While we experimented with two variants of BERT (small and large) and considered mean embeddings and concat (max + CLS token) embeddings, here we only report results for BERT large with mean embeddings, as it yielded the best results as seen in Table~\ref{tab:Interesults2}. All the Llama variants had very close performance metrics, whereas the OpenAI large text embedding model surpassed all of them. We attribute these small differences (\(\sim 2\%\) variation) in HitMusicLyricNet’s performance to variations in each model’s training data and architecture, since none was specifically trained for our downstream task, leading to large differences in rich embedding representation.

Hence with HitMusicLyricNet, we used the OpenAI large text embeddings and the SELU activation with MSE loss function in lyricsAENet. Overall, we achieved close to a 9\% improvement compared to the SOTA architecture. Dropping the lyrics feature pipeline and retraining and testing HitMusicLyricNet led to performance metrics comparable to that of HitMusicNet, validating the effectiveness of our proposed lyric feature extraction pipeline using LLMs and the overall enhancements in the music popularity prediction pipeline. A detailed ablation study for each feature set is provided in Appendix \ref{sec:appendixA}.

\section{Error Analysis}\label{error}
While HitMusicLyricNet surpasses the state-of-the-art baseline, an in-depth error analysis is necessary for real-world applications and future enhancements. In this section, we examine global residual errors, assess feature interpretability and impact via SHAP and LIME, and analyze social metadata to uncover any systematic biases and error patterns. All analyses are performed using the test set.

\begin{figure}[h]
\includegraphics[width=0.98\linewidth]{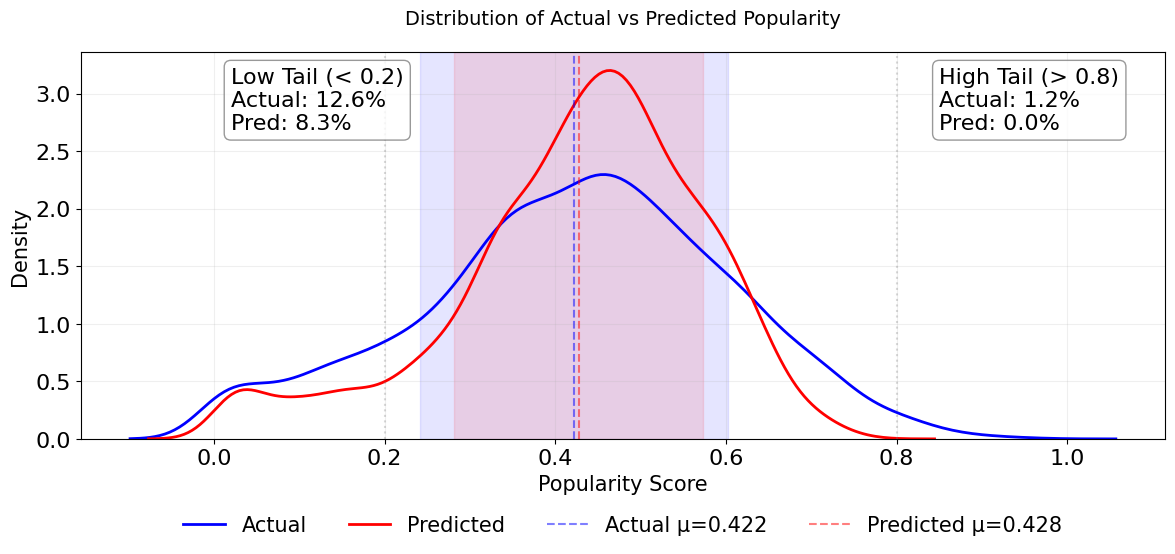} \hfill
  \caption {\label{fig:pop_dist} Actual (blue) vs. predicted (red) music popularity distributions on test set, showing prediction compression at both tails with aligned means ($\mu_{\text{actual}}=0.422$, $\mu_{\text{predicted}}=0.428$).}
  \vspace{-0.35cm}
\end{figure} 

\subsection{Global Residual Error Analysis}
Figure \ref{fig:pop_dist} compares the actual and predicted music popularity distributions. Although the means are nearly identical ($\mu_{\text{actual}} = 0.422$, $\mu_{\text{predicted}} = 0.428$), the predicted distribution's tails are compressed. The model predicts only 8.3\% of songs with popularity below 0.2 (compared to 12.6\% in the actual data) and fails to predict any songs with popularity above 0.8 (versus 1.2\% in the actual data). This regression towards the mean reflects both the limited representation of extreme popularity cases in SPD dataset and also the model's particular difficulty in capturing patterns of highly popular songs. 

\begin{figure}[h]
\includegraphics[width=0.98\linewidth]{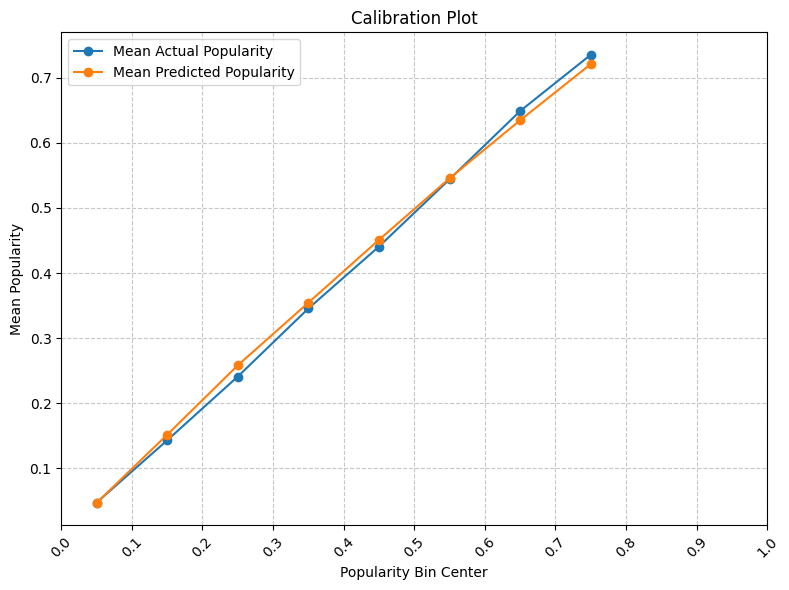} \hfill
  \caption {\label{fig:residual_caliberation_plot} Model calibration plot showing alignment between mean predicted and actual popularity per bin.}
  \vspace{-0.2cm}
\end{figure}

The calibration plot (Fig. \ref{fig:residual_caliberation_plot}) also indicates a strong alignment between predicted and actual music popularity within most bins, with the highest precision in the 0.4-0.6 range where data density peaks.

\subsection{Interpretability Analysis}
To understand the overall impact of non-interpretable latent representation of music audio and lyrics and the explicit metadata, we used SHAP (SHapley Additive exPlanations)\cite{10.5555/3295222.3295230}, and LIME (Local Interpretable Model-agnostic Explanations) \cite{ribeiro2016whyitrustyou} techniques on a randomly sampled 10\% of test data. 

\begin{figure}[h]
\includegraphics[width=0.98\linewidth]{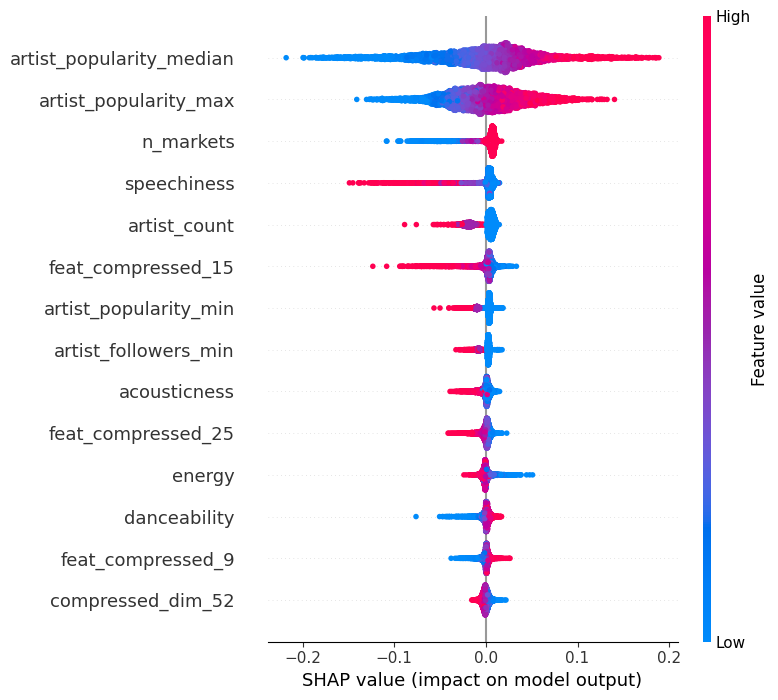} \hfill
  \caption {\label{fig:shape_global_analysis} SHAP value distributions for top 15 features across all modalities, with artist-related features showing highest impact on model predictions.}
\end{figure}

On analyzing the outcome of SHAP (Fig \ref{fig:shape_global_analysis}), artist popularity was the strongest predictor of music popularity with SHAP values ranging from $-0.2$ to $+0.2$. The compressed audio features showed a decreasing impact across sequential layers, indicating that earlier layers captured more predictive patterns. Lyric embeddings showed a moderate but consistent impact unless there is a significant deviation  from the typical pattern. LIME analysis supported these findings and substantiated detailed insights on decision boundaries within feature values as presented in Appendix \ref{sec:appendixB}.

\begin{figure}[h]
\includegraphics[width=0.98\linewidth]{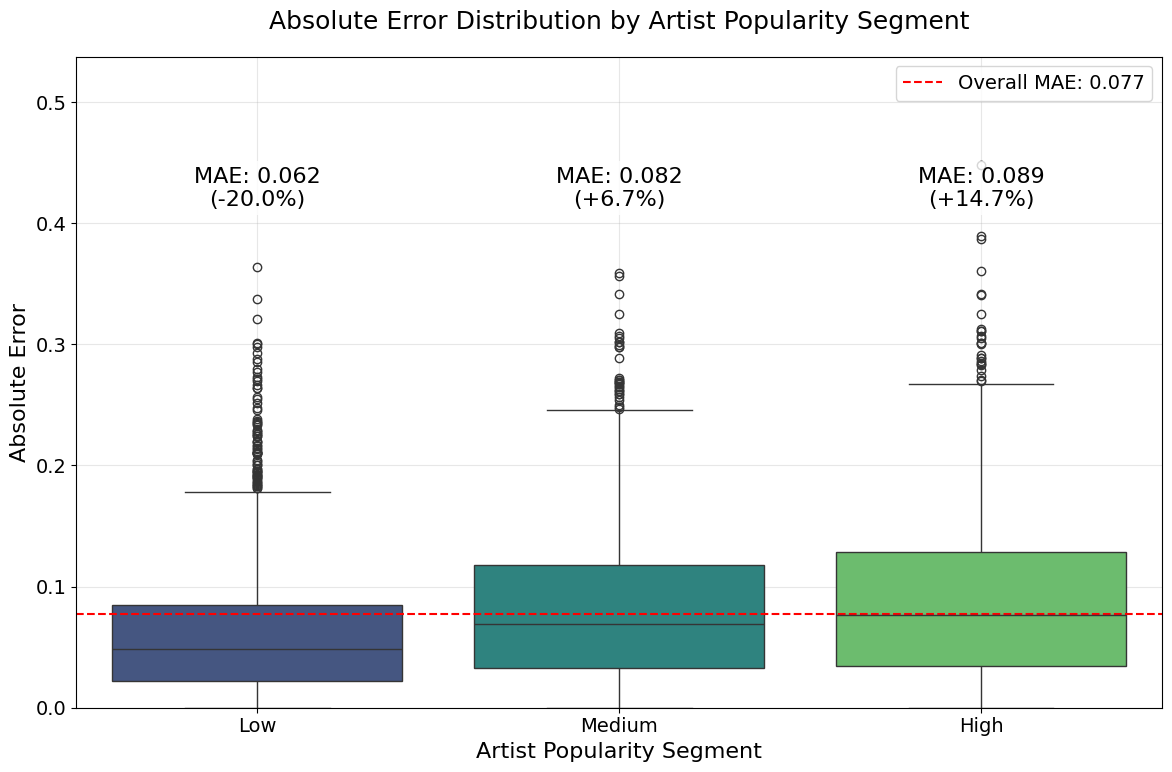} \hfill
  \caption {\label{fig:error_artist_popularity} Error distribution across artist popularity segments, showing MAE increase from low ($\mu=0.062$) to high ($\mu=0.089$) versus overall MAE ($\mu=0.077$).}
  \vspace{-0.2cm}
\end{figure} 

\subsection{Metadata and Artist-Level Analysis}
In the previous section, we observed that artist popularity is a dominant predictor of song popularity. To assess its impact and bias, we segmented the test set into three groups (low, medium, and high) based on artist popularity using quantiles. As shown in Fig.\ref{fig:error_artist_popularity}, songs composed by artists with low popularity have an MAE 20\% below the global MAE, while those in the medium and high segments exhibit MAEs 6.7\% and 14.7\% above it, respectively. Furthermore, LIME analysis (appendix \ref{sec:appendixB}) identified decision boundaries for artist popularity were at 0.19 and 0.39. Combined with the challenge of predicting the extreme right tail (Fig. \ref{fig:pop_dist}), these findings indicate that while artist popularity is a strong predictor for low- and mid-popularity songs, it falls short for highly popular tracks. Therefore, identifying patterns and strong predictors for highly popular songs still remains a research challenge.

Additionally, a year-wise error analysis (Fig. \ref{fig:error_release_year}) shows that both MAE and its variance were significantly higher in the 1990s and early 2000s. Since 2005, however, errors have stabilized—likely reflecting a training bias towards recent years and also aligning with Spotify’s song popularity score calculation, which emphasizes more on recent time metrics.

\section{Conclusion and Future Work}
\label{sec:bibtex}
The work presented in this paper showcases the power of leveraging lyrics to predict the popularity of a song, with the help of LLMs with capabilities of capturing the deeper meaning of sentences using embeddings. We believe that advancements in music-aware language models will lead to more explainable and expressive lyric features based on domain-specific knowledge. This research presented a novel architecture, HitMusicLyricNet, along with an ablation study. HitMusicLyricNet beats the SOTA by 9\% by incorporating lyric embeddings and improving upon the SOTA architecture. With advancements in compression techniques and multimodal learning architecture, we believe accuracy and commercial use can be improved. Furthermore, with advancements in audio representation learning using neural audio codecs, richer music audio representations can be scoped into the study. Current research aggregates features over an entire song. However, contemporary phenomena of virality suggest that local features within different musical segments need to be studied deeply and cannot be ignored given the micro-content consumption driven by platforms like Instagram and SnapChat.


\section{Limitation}
Our findings may be constrained by genre, demographic, and cultural variability not fully captured in the current experimental setup. While LLMs such as BERT and LLaMA-3 enable deeper semantic modeling of lyrics, their general-purpose training limits their ability to capture music-specific linguistic patterns. Despite careful regularization, the high dimensionality of lyric embeddings presents inherent risks of overfitting. Moreover, as these embeddings are evaluated solely through downstream task performance, their intrinsic quality in representing lyrical content remains underexplored. Finally, the opacity of these feature vectors limits interpretability, pointing to a need for more explainable models of lyric representation.



\bibliography{custom}

\appendix

\section{Ablation Study}\label{sec:appendixA}


In this section, we study how different modalities contribute to our model's music popularity predictive strength. Table~\ref{tab:Modality_1} shows model performance for each combination of our four feature types: high-level audio (HH), low-level audio (LL), lyrics embeddings (LR), and metadata (M).

The model works best when it uses all modalities, with a test MAE of $0.0772$. If we exclude lyrics embeddings, the test MAE increases by $10.4\%$ to $0.0852$, highlighting the usefulness of our proposed lyrics feature pipeline. Notably, using only high-level features and metadata along with lyrics (HH, LR, M) gives comparable performance to using all the modalities features, indicating some redundancy in low-level audio features. The role of social context is apparent when we strip metadata by utilizing only audio and lyrics features (HH, LL, LR), which makes the test MAE rise by $40.2\%$ to $0.1082$. Performance suffers most significantly if we use only audio features (HH, LL) and obtain a test MAE of $0.1196$.

\begin{table}[h]
    \centering
        \small
        \begin{tabular}{>{\centering\arraybackslash}p{2.2cm}>{\centering\arraybackslash}p{1.3cm}>{\centering\arraybackslash}p{1.3cm}>{\centering\arraybackslash}p{1.3cm}}
            \hline
            \textbf{Modality Config} & \textbf{MAE (Train)} & \textbf{MAE (Val)} & \textbf{MAE (Test)} \\
            \hline
            \textit{HH, LL, LR, M}  & \textbf{0.0761} & \textbf{0.0743} & \textbf{0.0772} \\
            \textit{HH, LL, M} & 0.0818 & 0.0841 & 0.0852 \\
            \textit{HH, LL, LR} & 0.1059 & 0.1037 & 0.1082 \\
            \textit{HH, LR, M} & 0.0767 & 0.0765 & 0.0795 \\
            \textit{HH, LL} & 0.1188 & 0.1175 & 0.1196 \\
            \textit{LR, M} & 0.0810 & 0.0811 & 0.0805 \\
            \hline
        \end{tabular}
        \caption{\label{tab:Modality_1} Results of training and testing HitMusicLyricNet with different modality combinations. HH: High-level audio features, LL: Low-level audio features, LR: Lyrics embeddings features, M: Metadata features.}
\end{table}

To further understand individual modality performance, we conducted isolated training experiments as shown in Table \ref{tab:Modality_2}. Single-modality tests ascertain that metadata features (M) alone achieve the highest single-modality performance with a test MAE of 0.0968, verifying our initial observation about the importance of social context in music popularity prediction. Lyrics embeddings (LR) are similarly predictive to low-level audio features (LL), with test MAEs of 0.1193 and 0.1229, respectively. High-level audio features (HH) are slightly worse in isolation with a test MAE of 0.1272. These results show that while each modality contains valuable information, their combination creates synergistic effects that significantly improve prediction accuracy, as evidenced by the better performance of the full model in Table \ref{tab:Modality_1}.

\begin{table}[h]
    \centering
        \small
        \begin{tabular}{>{\centering\arraybackslash}p{2.2cm}>{\centering\arraybackslash}p{1.3cm}>{\centering\arraybackslash}p{1.3cm}>{\centering\arraybackslash}p{1.3cm}}
            \hline
            \textbf{Modality Config} & \textbf{MAE (Train)} & \textbf{MAE (Val)} & \textbf{MAE (Test)} \\
            \hline
            \textit{LL}  & 0.1234 & 0.1218 & 0.1229 \\
            \textit{HH} & 0.1260 & 0.1266 & 0.1272 \\
            \textit{LR} & 0.1208 & 0.1189 & 0.1193 \\
            \textit{M} & 0.1026 & 0.0956 & 0.0968 \\
            \hline
        \end{tabular}
        \caption{\label{tab:Modality_2} Performance comparison of individual modalities in predicting song popularity, showing the relative strength of each feature type in isolation.}
        \vspace{-0.3cm}
\end{table}

\section{Error and Feature Importance Analysis} \label{sec:appendixB}

To supplement our error analysis discussed in Section \ref{error}, we conducted a detailed investigation of model behavior through two complementary approaches: (1) analysis of prediction residuals and their distribution patterns, and (2) assessment of feature importance across different modalities using SHAP and LIME techniques. 

\begin{figure}[h]
\includegraphics[width=0.98\linewidth]{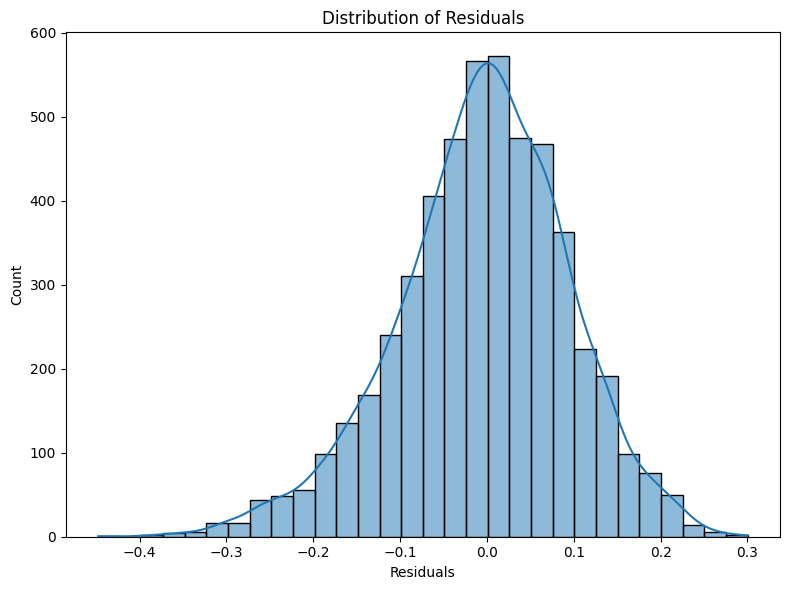} \hfill
  \caption {\label{fig:residual_distribution} Distribution of prediction residuals centered at $\mu \approx 0.0$, showing approximately normal spread with slight negative skewness.}.
  \vspace{-0.2cm}
\end{figure} 

\begin{figure}[h]
\includegraphics[width=0.98\linewidth]{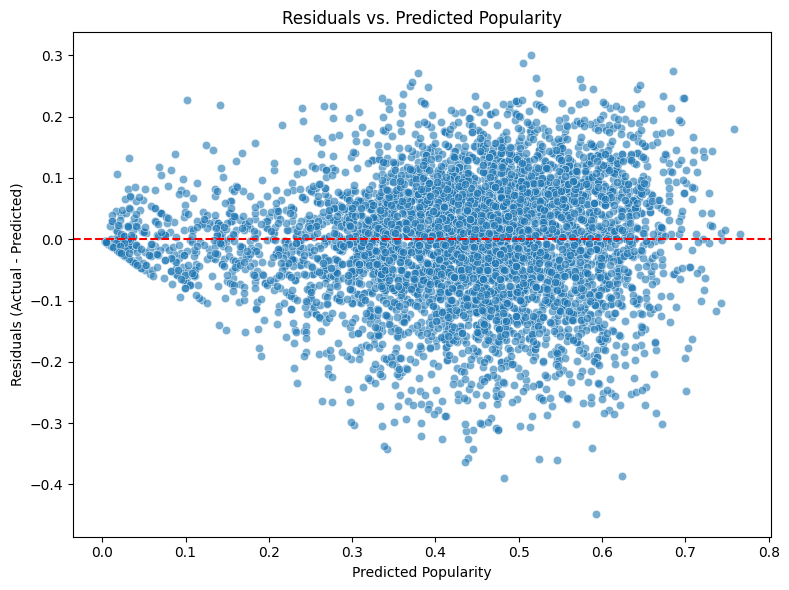} \hfill
  \caption {\label{fig:residual_scatter_plot} Scatter plot of residuals vs predicted popularity values showing error distribution across popularity ranges.}
  \vspace{-0.1cm}
\end{figure}

Analysis of the residual distribution (Figure \ref{fig:residual_distribution}) shows a quasi-normal pattern centered at zero, with about 95\% of forecasts falling within ±0.2 of actual values. The distribution shows minimal negative skewness, suggesting a small inclination toward underestimating in extreme conditions. With variance amplification in the mid-popularity range (0.3–0.6) and more limited errors at the extremes, the residual scatter plot against predicted popularity (Figure \ref{fig:residual_scatter_plot}) shows heteroscedastic behavior.

\begin{figure}[h]
\includegraphics[width=0.98\linewidth]{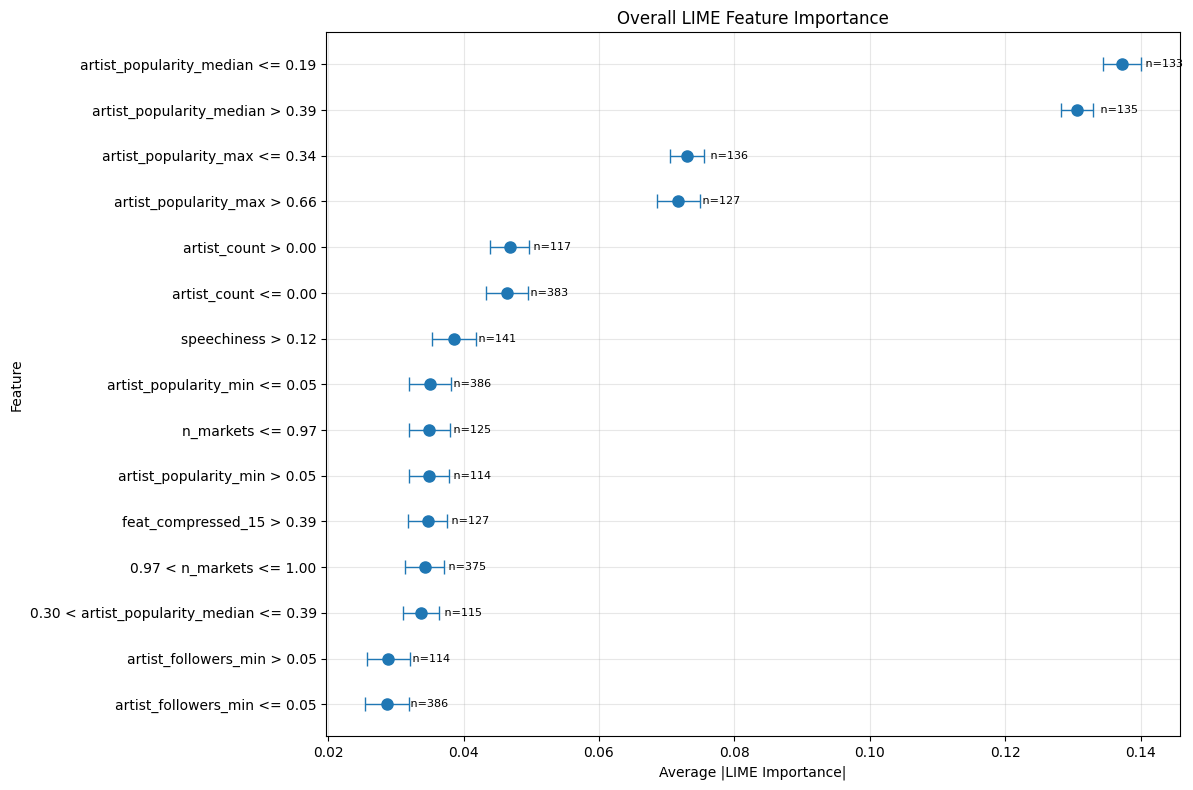} \hfill
  \caption {\label{fig:lime_aggregate_imp} Aggregated global LIME feature importance scores across the test set, demonstrating artist popularity thresholds as dominant predictors. Values represent absolute LIME coefficients with 95\% confidence intervals, $n$ indicates per-feature sample size.}
  \vspace{-0.3cm}
\end{figure} 

The LIME study shows varied trends in feature relevance over multiple modalities. With artist popularity thresholds ($\leq 0.19$ and $>0.39$) displaying the highest importance scores ($\sim$0.13), artist-related metadata dominates the prediction process in the general feature landscape (Figure \ref{fig:lime_aggregate_imp}). This division implies that the algorithm has learnt different behavioral patterns for artists at various degrees of popularity.

\begin{figure}[h]
\includegraphics[width=0.98\linewidth]{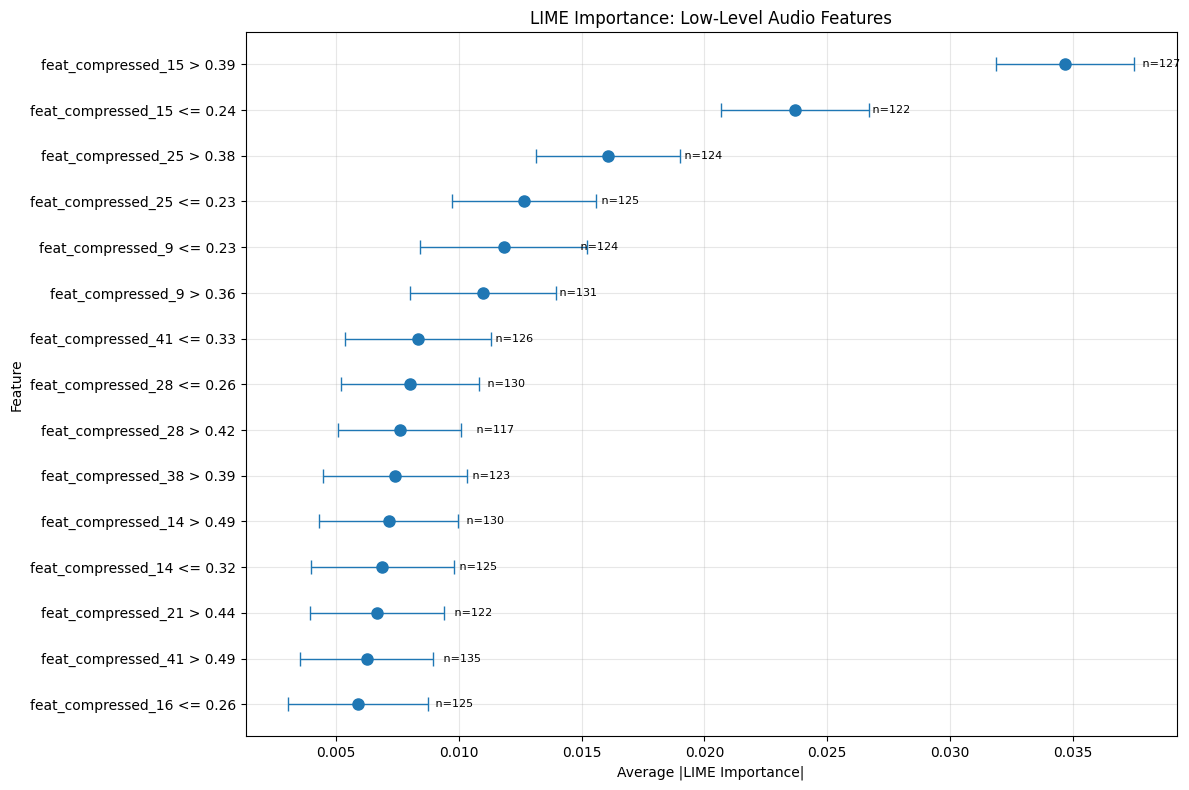} \hfill
  \caption {\label{fig:lime_ll} LIME importance scores for compressed low-level audio features, showing early compressed dimensions (particularly feat\_compressed\_15) having higher predictive power.}
  \vspace{-0.1cm}
\end{figure}

Early compressed dimensions (especially \texttt{feat\_compressed\_15}) have higher predictive weight than later ones, therefore displaying a hierarchical importance structure in the low-level audio characteristics (Figure \ref{fig:lime_ll}). This trend shows that in its first compression layers, our AudioAENet efficiently retains fundamental acoustic information.

A deeper interpretation of the LIME results for lyric-embedding characteristics shows that although some compressed dimensions (such as 52 and 54) often show themselves as most essential, their impact on the prediction is not consistent across all samples. Particularly several threshold splits for these dimensions (e.g., \texttt{compressed\_dim\_52} $> 0.05$ vs. $\leq 0.03$) point to a non-linear or boundary-based relationship: the model may be using these latent factors to distinguish between songs that surpass certain ``lyrical thresholds'' (perhaps tied to vocabulary, theme, or semantic content) and those that do not.

\begin{figure}[h]
\includegraphics[width=0.98\linewidth]{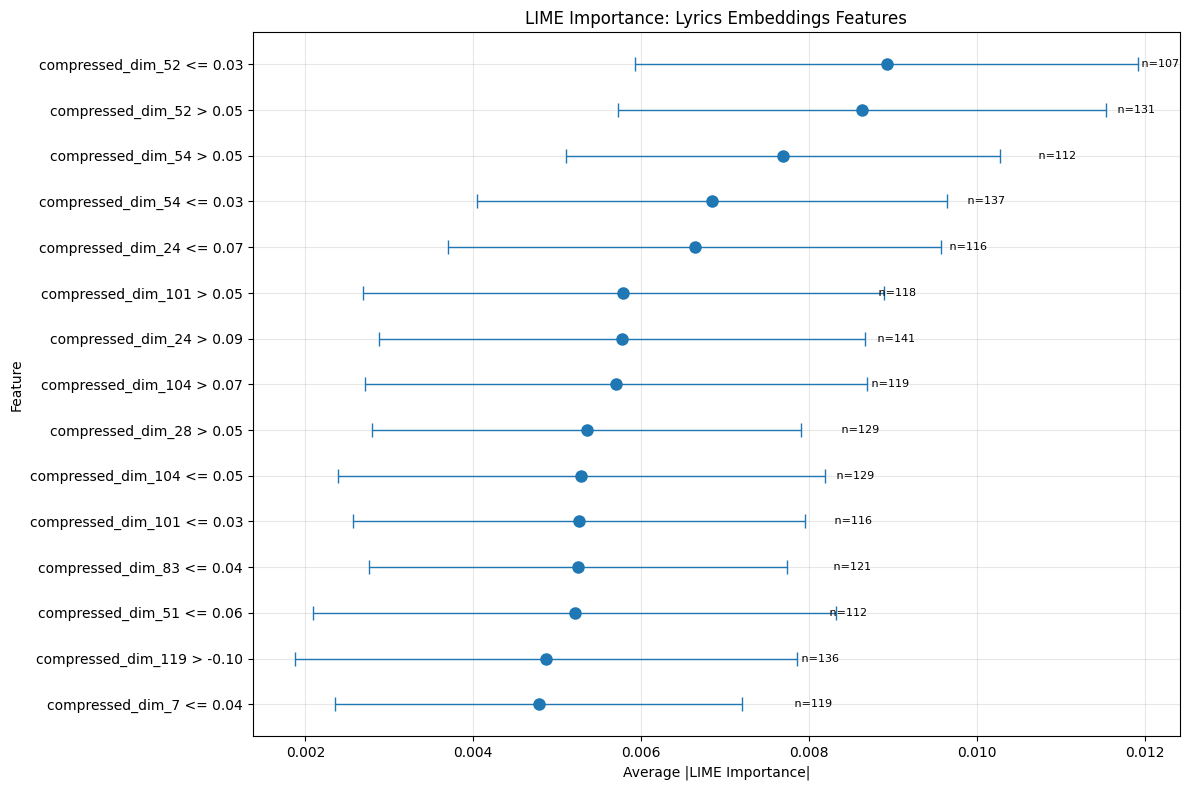} \hfill
  \caption {\label{fig:lime_lyric} LIME importance scores for compressed lyric embedding dimensions, highlighting threshold-based importance patterns in dimensions 52 and 54. Wider confidence intervals indicate more variable impact of lyrical features.}
  \vspace{-0.3cm}
\end{figure} 

\begin{figure}[h]
\includegraphics[width=0.98\linewidth]{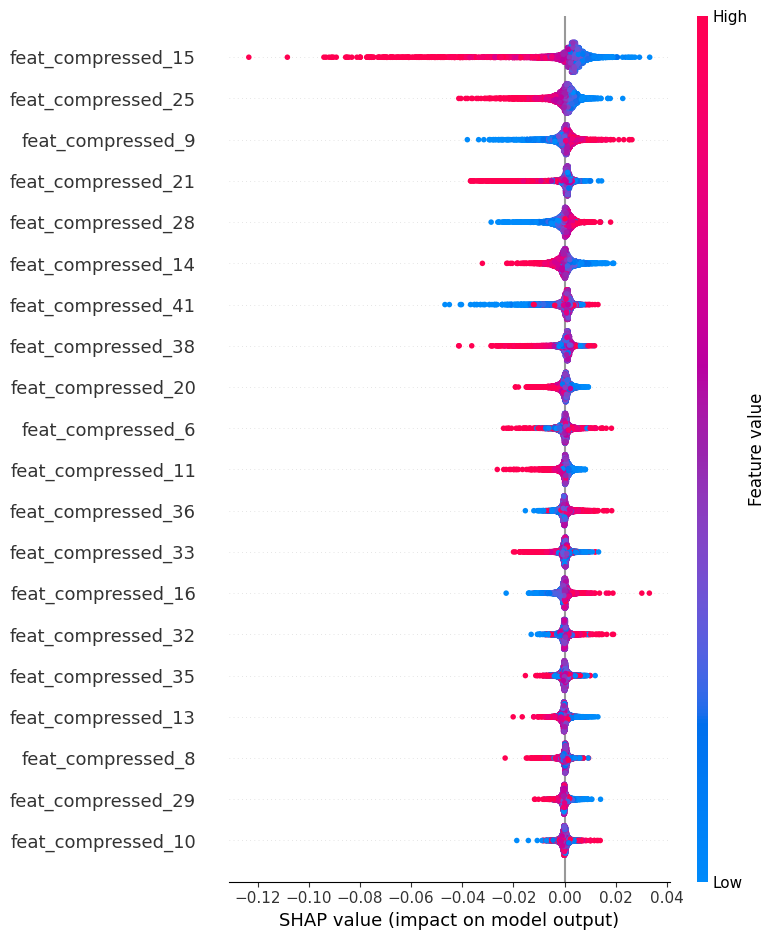} \hfill
  \caption {\label{fig:shap_ll} SHAP values for compressed audio features, showing stronger impact of early dimensions (feat\_compressed\_15) with values ranging from $-0.12$ to $+0.04$. Color indicates original feature value magnitude (blue=low, red=high).}
  \vspace{-0.1cm}
\end{figure} 

\begin{figure}[h]
\includegraphics[width=0.98\linewidth]{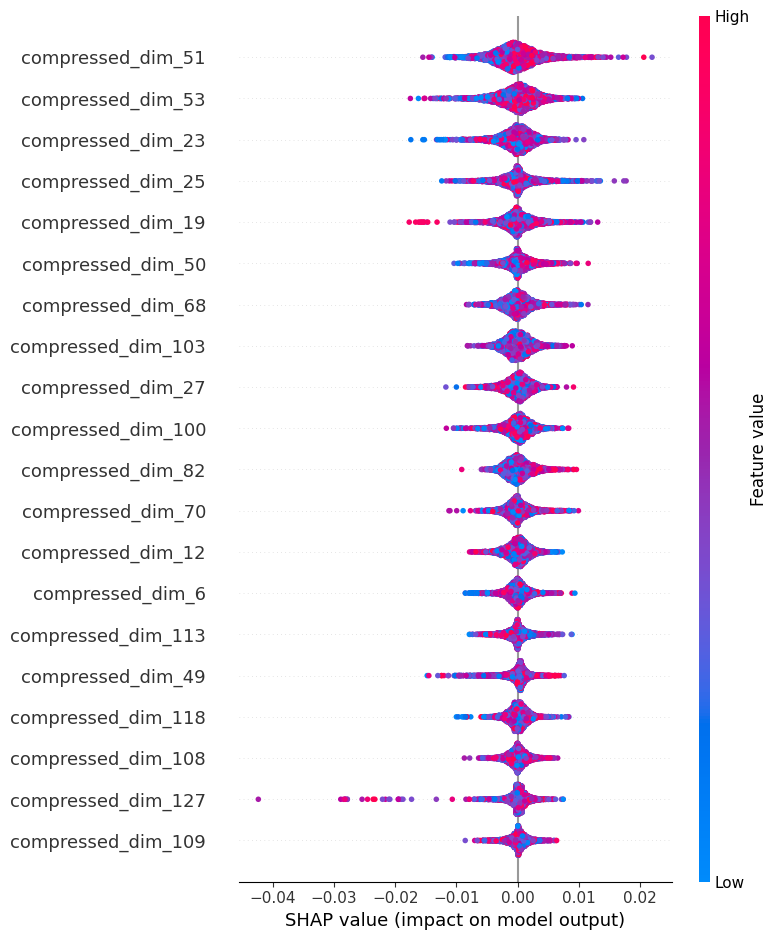} \hfill
  \caption {\label{fig:shap_lyric} SHAP values for lyric embedding dimensions, revealing more symmetric distributions around zero ($\pm0.02$) with notable outliers in dim\_127. Colors represent embedding magnitude in each dimension.}
\end{figure}

The SHAP analysis shows complex patterns in how lyrical elements influence popularity predictions (Figures 13--14).  For lyrics (Figure \ref{fig:shap_lyric}), while most dimensions cluster tightly around zero ($\pm 0.01$ SHAP value), several dimensions demonstrate different patterns.  The top dimensions (51--25) show bigger influence distributions and more extreme outlier points.  Particularly in dimensions 51, 53, and 23, an interesting trend in the color distribution shows that positive SHAP values often correspond with greater feature values (red) and negative with lower values (blue). This implies that these measures reflect poetic aspects that, either highly present or missing, always affect popularity in particular directions. With scarce but considerable negative effects (reaching $-0.04$) and a mixed color distribution, \texttt{Compressed\_dim\_127} exhibits a distinctive pattern that indicates it captures complicated lyrical features that influence popularity irrespective of their size.

By contrast, the audio features (Figure \ref{fig:shap_ll}) exhibit more asymmetric impact distributions, especially in \texttt{feat\_compressed\_15} with the highest magnitude of impact ($-0.12$ to $0.04$). Early compressed audio characteristics (15, 25, 9) show significantly higher SHAP values than later dimensions, therefore confirming the capacity of our autoencoder to retain important acoustic information in its first layers. Notably, while audio features tend to have larger absolute SHAP values than lyrics features, they also show more defined directionality in their effects, suggesting more deterministic relationships with popularity predictions.

\begin{figure}[h]
\includegraphics[width=0.98\linewidth]{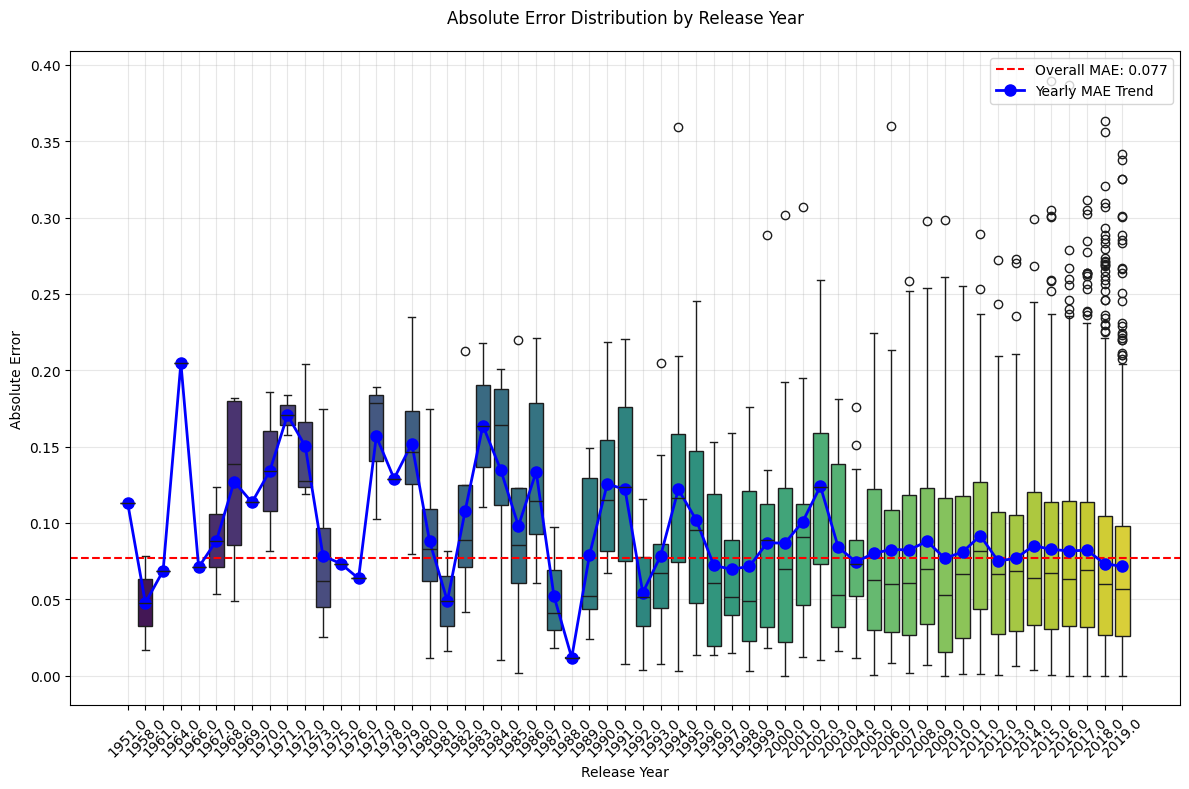} \hfill
  \caption {\label{fig:error_release_year} Year-wise absolute error distribution (1950--2019) showing higher error variance in early decades (1990s) followed by stabilization post-2005. Box plots show error distributions per year, blue line tracks yearly MAE trend, and red dashed line indicates overall MAE of 0.077 .}
  \vspace{-0.3cm}
\end{figure} 
\end{document}